\documentclass[12pt, a4paper]{article}
\usepackage[margin=2.5cm]{geometry}
\usepackage{amsthm, amsmath, amssymb, bm, bbm, ascmac, listings, algorithm, algpseudocode, tikz}
\usepackage[backend=biber, style=apa, citestyle=authoryear]{biblatex}
\renewbibmacro*{cite}{%
  \printnames{labelname}%
  \addspace
  \printtext[parens]{\printfield{year}}}
\usetikzlibrary{fit}
\theoremstyle{definition}
\newtheorem{dfn}{Definition}[section]
\newtheorem{prop}[dfn]{Proposition}
\newtheorem{lem}[dfn]{Lemma}
\newtheorem{thm}[dfn]{Theorem}

\title{Efficiency in the Roommates Problem\footnote{We would like to express our sincere gratitude to Professor Michihiro Kandori and Professor Fuhito Kojima for their invaluable guidance and insightful comments. We are also deeply grateful to Professor David Manlove for kindly introducing us to related research, which provided invaluable insights.}
}
\author{
Keita Kuwahara\thanks{The Faculty of Economics, the University of Tokyo, Japan. Email: kuke0303@g.ecc.u-tokyo.ac.jp}
}

\date{\today}

\begin{document}

\maketitle

\begin{abstract}
We propose an $O(n^2)$-time algorithm to determine whether a given matching is efficient in the roommates problem.
\end{abstract}

\section{Introduction}
Matching theory is an important field of research in economics, computer science, and mathematics, with applications spanning education, healthcare, and labor markets. Among these, the roommates problem, introduced by \cite{gale1962college}, stands out as one of the classic challenges in matching theory. It involves pairing individuals within a single group based on mutual evaluations and is recognized as a generalization of the stable marriage problem. In this paper, we investigate methods for determining whether a given matching is efficient in the roommates problem.

To assess the efficiency of a matching, it is necessary to verify that no other matching Pareto dominates it. However, because the number of possible matchings grows exponentially with the number of agents $n$, any algorithm that checks each matching individually cannot operate in polynomial time. \cite{aziz2013pareto} proposed an algorithm with a computational complexity of $O(n^3)$ to address this challenge. On the other hand, \cite{morrill2010roommates} introduced an algorithm that runs in $O(n^2)$ time under the assumption that every agent is matched with someone. In this paper, we extend the algorithm of \cite{morrill2010roommates} and propose an algorithm that maintains an $O(n^2)$ time complexity even in models where agents are allowed to remain unmatched.

\cite{abraham2004pareto} introduced an algorithm with a time complexity of \(O(m)\), where \(m\) represents the total number of agents that are considered acceptable as potential matches, summed across all agents. The maximum possible value of \(m\) is \(n(n-1)\). 
The key difference between \cite{abraham2004pareto} and our work is that while \cite{abraham2004pareto} focuses on matchings between mutually acceptable agents, we explore an algorithm to determine whether a given matching is efficient even if some agents are matched with unacceptable partners.
Our model is designed as a generalization of their model.

\section{Preliminaries}\label{sec:preliminaries}
For any integer $n \geq 3$, let $N = \{1, 2, \dots, n\}$ be a finite set of agents. Denote by $\mathcal{P}$ the set of linear orders or strict preferences on $N$. A function $\mu : N \rightarrow N$ represents a \emph{matching} if $\mu(\mu(i)) = i$ for every $i \in N$.
Note that $\mu(i)=j$ denotes that agent $i$ is matched with agent $j$, while $\mu(i)=i$ denotes that agent $i$ is not matched with anyone.
Let $\mathcal{M}$ be the set of matchings.
For distinct matchings $\mu, \mu' \in \mathcal{M}$, if $\mu(i) \succeq_i \mu'(i)$ for all $i \in N$, then we say that $\mu$ \emph{Pareto dominates} $\mu'$. A matching $\mu \in \mathcal{M}$ is \emph{efficient} if there is no matching that Pareto dominates it. \par

As in \cite{morrill2010roommates}, given a matching $\mu \in \mathcal{M}$ and a preference profile $\succ \in \mathcal{P}^n$, we define an undirected graph $G$ as follows:
\begin{itemize}
    \item Each vertex represents an agent. There are $n$ vertices in total.
    \item Special edges connect agents that are matched to each other; that is, each agent is connected by a special edge to its matched partner. If agent $i$ is not matched with anyone, a self-loop connecting $i$ to itself is formed.
    \item normal edges connect every pair $(i,j)\in N\times N$ satisfying
    \[
    j \succ_i \mu(i) \quad \text{and} \quad i \succ_j \mu(j).
    \]
    In this case, for every agent $i$ satisfying  
    \[
        i \succ_i \mu(i),
    \]
    a normal edge is also added, forming a self-loop that connects $i$ to itself.
\end{itemize}\par

\begin{figure}[ht]
    \centering
    \begin{minipage}{0.3\textwidth}
        \centering
        \begin{tikzpicture}
            \node[circle, draw] (4) at (1.5,1.5) {4};
            \node[circle, draw] (1) at (0.5,0.5) {1};
            \node[circle, draw] (8) at (-0.5,-0.5) {8};
            \node[circle, draw] (5) at (-1.5,-1.5) {5};
            
            \draw[double=black, double distance=2pt] (4) -- (1);
            \draw[double=black, double distance=2pt] (8) -- (5);
            \draw[double distance=2pt] (4) to[out=90,in=0,looseness=5] (4);
            \draw[double distance=2pt] (5) to[out=180,in=270,looseness=5] (5);
            \draw[double distance=2pt] (8) -- (1);
        \end{tikzpicture}
    \end{minipage}
    \hspace{0.1\textwidth}
    \begin{minipage}{0.3\textwidth}
        \centering
        \begin{tikzpicture}
            \node[circle, draw] (1) at (0,2) {1};
            \node[circle, draw] (3) at (1.732,1) {3};
            \node[circle, draw] (9) at (1.732,-1) {9};
            \node[circle, draw] (2) at (0,-2) {2};
            \node[circle, draw] (7) at (-1.732,-1) {7};
            \node[circle, draw] (6) at (-1.732,1) {6};

            \draw[double=black, double distance=2pt] (3) -- (9);
            \draw[double=black, double distance=2pt] (2) -- (7);
            \draw[double=black, double distance=2pt] (6) -- (1);
            
            \draw[double distance=2pt] (1) -- (3);
            \draw[double distance=2pt] (9) -- (2);
            \draw[double distance=2pt] (7) -- (6);
        \end{tikzpicture}
    \end{minipage}
    \caption{The figure on the left illustrates an alternating path, whereas the figure on the right illustrates an alternating cycle. Special edges are depicted in black, and normal edges are depicted in white.}
    \label{fig:alternating structures}
\end{figure}
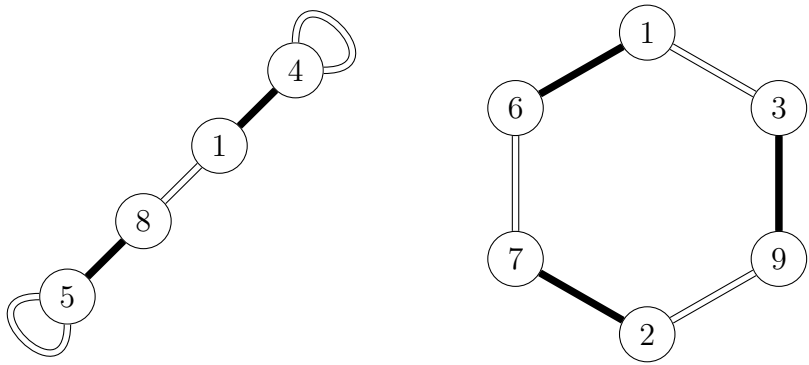

An \emph{alternating path} is defined as a path in which special and normal edges appear alternately, which is not a cycle, and whose terminal vertices have self-loops. Conversely, an \emph{alternating cycle} is defined as a cycle in which special and normal edges appear alternately.
Figure \ref{fig:alternating structures} shows examples of an alternating path and an alternating cycle.

\section{Algorithm}\label{sec:algorithm}
First, we establish the following lemma regarding the relationship between the graph and efficiency.

\lem\label{lem:graph and efficiency}
A matching satisfies efficiency if and only if its graph contains neither alternating paths nor alternating cycles.
\proof
We first prove the "only if" direction: if the graph $G$ contains an alternating path or an alternating cycle, then the matching $\mu$ does not satisfy efficiency.
We define a different matching $\mu'$ as follows:
\begin{enumerate}
    \item Arbitrarily select an alternating path or an alternating cycle in $G$.
    \item For every normal edge contained within the selected structure, match the agents connected by that edge (including cases where an agent is matched to themselves).
    \item All other agents who are not yet matched are assigned the same partners as in $\mu$.
\end{enumerate}
Note that when an agent is matched via a normal edge, their original partner in $\mu$ is also reassigned through another normal edge.
Thus, $\mu'$ constitutes a valid matching. Since $\mu'$ Pareto dominates $\mu$, the matching $\mu$ is not efficient.\par

Next, we prove the "if" direction: if the matching $\mu$ does not satisfy efficiency, then the graph $G$ contains an alternating path or an alternating cycle.
Let $\mu'$ be any matching that Pareto dominates $\mu$.\par

First, consider the case where there exists some $i \in N$ such that $\mu(i) = i$ but $\mu'(i) \neq i$.
Define $j_1 = \mu'(i)$. 
For $k = 1,3,5,\dots$, define the sequence inductively as follows:
\begin{itemize}
    \item $j_{k+1} = \mu(j_k),$
    \item $j_{k+2} = \mu'(j_{k+1})$.
\end{itemize}
Let $k^*$ be the smallest $k$ such that $j_k = j_{k+1}$. 
Since the set of agents is finite, such a $k^*$ necessarily exists.
Any agent $j$ whose matching partner differs between $\mu$ and $\mu'$ satisfies $\mu'(j) \succ_j \mu(j)$, meaning that $j$ and $\mu'(j)$ are connected by a normal edge.
Therefore, the path passing through $i, j_1, j_2, \dots, j_{k^*}$ forms an alternating path.\par

Similarly, consider the case where there exists some $i \in N$ such that $\mu'(i) = i$ but $\mu(i) \neq i$.
Define $j_1 = \mu(i)$. 
For $k = 1,3,5,\dots$, define the sequence inductively as follows:
\begin{itemize}
    \item $j_{k+1} = \mu'(j_k),$
    \item $j_{k+2} = \mu(j_{k+1})$.
\end{itemize}
Let $k^*$ be the smallest $k$ such that $j_k = j_{k+1}$.
Note that $i$ and $\mu'(i)$, as well as any $j_k$ and $\mu'(j_k)$, are connected by a normal edge.
Thus, the path passing through $i, j_1, j_2, \dots, j_{k^*}$ forms an alternating path.\par

Finally, consider the case where, for all $i \in N$ such that $\mu(i) \neq \mu'(i)$, it holds that $\mu(i) \neq i$ and $\mu'(i) \neq i$.
Fix any agent $i \in N$ whose matching partner differs between $\mu$ and $\mu'$.
Define $j_1 = \mu(i)$. 
For $k = 1,3,5,\dots$, define the sequence inductively as follows:
\begin{itemize}
    \item $j_{k+1} = \mu'(j_k),$
    \item $j_{k+2} = \mu(j_{k+1})$.
\end{itemize}
Let $k^*$ be the smallest $k$ such that $j_k = i$.
By assumption, we have $\mu(j_k) \neq \mu'(j_k)$, $\mu(j_k) \neq j_k$, and $\mu'(j_k) \neq j_k$ for any $k$.
Since the set of agents is finite, such a $k^*$ necessarily exists.
Note that any $j_k$ and $\mu'(j_k)$ are connected by a normal edge.
Therefore, the cycle passing through $i, j_1, j_2, \dots, j_{k^*}$ forms an alternating cycle.\par

To summarize the above discussion, it follows that $G$ contains an alternating path or an alternating cycle.
\qed\\[1ex]\par

Given a graph $G$, we define a \emph{modified graph} $G'$ as follows:
\begin{enumerate}
    \item For any vertex $i\in N$ with a self-looping special edge in $G$, introduce a new vertex $i+n$ (referred to as a \emph{virtual vertex}) and connect $i$ and $i+n$ with a special edge.
    \item For any two distinct vertices $i$ and $j$ that are not connected by an edge and satisfy one of the following conditions (either separately or together), connect $i$ and $j$ with a normal edge:
    \begin{itemize}
        \item It is a virtual vertex.
        \item It has a self-looping normal edge in $G$.
    \end{itemize}
    \item Remove all edges that form a self-loop (i.e., edges that connect an agent to itself).
\end{enumerate}

For a matching $\mu \in \mathcal{M}$, a distinct pair of agents $i, j \in N$ is an \emph{irrational pair} if $\mu(i) = j$, $i \succ_i j$, and $j \succ_j i$ hold simultaneously.
The relationship between the graph $G$ and the modified graph $G'$ is described as follows.

\lem\label{lem:graph and modified graph}
Suppose that a matching $\mu$ contains no irrational pairs. Then, its graph contains an alternating path or an alternating cycle if and only if its modified graph contains an alternating cycle.
\proof
We first prove the "only if" direction: if the graph $G$ contains an alternating path or an alternating cycle, then the modified graph $G'$ contains an alternating cycle.
If $G$ contains an alternating cycle, it is evident that $G'$ also contains an alternating cycle. Therefore, we assume that $G$ does not contain an alternating cycle but does contain an alternating path.
Fix an arbitrary alternating path $P$ in $G$. Without loss of generality, let the two endpoints of $P$ be labeled as $1,2 \in N$.\par

First, consider the case where $\mu(1) = 1$ and $\mu(2) = 2$. In this case, note that the alternating path $P$ from $1$ to $2$ starts with a normal edge and ends with a normal edge, except for the self-loops of its endpoints. Since both $1$ and $2$ have self-looping special edges in $G$, they are connected by special edges to $n+1$ and $n+2$ in $G'$, respectively. Since $n+1$ and $n+2$ are virtual vertices, they are connected by a normal edge in $G'$. Thus, the sequence $1, n+1, n+2, 2$, followed by the same path as $P$ back to $1$, constitutes an alternating cycle in $G'$.\par

Next, consider the case where $\mu(1) = 1$ and $\mu(2) \neq 2$. In this case, note that the alternating path $P$ from $1$ to $2$ starts with a normal edge and ends with a special edge, except for the self-loops of its endpoints. Since $1$ has a self-looping special edge in $G$, it is connected to $n+1$ by a special edge in $G'$. Also, since $2$ is an endpoint of $P$, it has a self-looping normal edge in $G$, meaning that $n+1$ and $2$ are connected by a normal edge in $G'$ by definition. Consequently, the sequence $1, i+1, 2$, followed by the same path as $P$ back to $1$, constitutes an alternating cycle in $G'$.
The case where $\mu(1) \neq 1$ and $\mu(2) = 2$ follows similarly by swapping $1$ and $2$.

Finally, consider the case where $\mu(1) \neq 1$ and $\mu(2) \neq 2$. In this case, note that the alternating path $P$ from $1$ to $2$ starts and ends with a special edge, except for the self-loops of its endpoints.
Since $\mu$ does not contain any irrational pairs, we have $\mu(1) \neq 2$. Also, since both $1$ and $2$ are endpoints of $P$, they have self-looping normal edges in $G$. Thus, $1$ and $2$ are connected by a normal edge in $G'$. Consequently, the sequence $1,2$, followed by the same path as $P$ back to $1$, constitutes an alternating cycle in $G'$.\par

From the above discussion, it follows that $G'$ contains an alternating cycle in all cases.\par

Next, we prove the "if" direction: if the modified graph $G'$ contains an alternating cycle, then the original graph $G$ contains either an alternating path or an alternating cycle.
Fix an alternating cycle $C$ in $G'$. 
To derive a contradiction, assume that $G$ contains neither an alternating path nor an alternating cycle. 
Thus, \( C \) must contain at least one edge that was added in the process of constructing \( G' \) from \( G \).\par

First, suppose that $C$ contains a virtual vertex. 
Select an arbitrary virtual vertex in $C$ and denote it as $i_1+n$ for some $i_1 \in N$. 
In $G$, it follows that $i_1$ has a self-looping special edge.
Therefore, $i_1$ does not have a self-looping normal edge in $G$.
Consequently, the vertex connected to $i_1$ by a normal edge in $C$, denoted as $i_2$, is also connected to $i_1$ by a normal edge in $G$.
Proceeding along $C$ in the order $i_1+n, i_1, i_2$, and continuing until returning to $i_1+n$, we define the sequence of vertices encountered as $i_3, i_4, \dots, i_k$, where $i_k = i_1+n$.
Now, let $j^*$ be the smallest index $j \geq 2$ such that the edge connecting $i_j$ and $i_{j-1}$ does not exist in $G$. 
Since the edge connecting $i_k$ and $i_{k-1}$ is absent in $G$, $j^*$ must exist.
Moreover, since the edge connecting $i_2$ and $i_1$ exists in $G$, we have $j^* \geq 3$.\par

First, suppose that $i_{j^*}$ is a virtual vertex. 
If $i_{j^*-1}$ is also a virtual vertex, then the edge connecting $i_{j^*-1}$ and $i_{j^*-2}$ does not exist in $G$, contradicting the definition of $j^*$. 
Thus, we have $i_{j^*-1} \in N$. 
If the edge connecting $i_{j^*}$ and $i_{j^*-1}$ in $G'$ is a normal edge, then in $G$, $i_{j^*-1}$ has a self-looping normal edge.
On the other hand, if the edge connecting $i_{j^*}$ and $i_{j^*-1}$ in $G'$ is a special edge, then in $G$, the special edge of $i_{j^*-1}$ is a self-loop.
In \( G \), the path that passes through \( i_1, i_2, \dots, i_{j^*-1} \) in the same way as \( C \) forms an alternating path in either case, leading to a contradiction.\par

Next, suppose that $i_{j^*}$ is not a virtual vertex. 
As before, if $i_{j^*-1}$ were a virtual vertex, then the edge connecting $i_{j^*-1}$ and $i_{j^*-2}$ would not exist in $G$, contradicting the definition of $j^*$.
Thus, we have $i_{j^*-1} \in N$. 
It follows that in $G$, $i_{j^*-1}$ has a self-looping normal edge by definition.
This again implies that in \( G \), the path that passes through \( i_1, i_2, \dots, i_{j^*-1} \) in the same way as \( C \) forms an alternating path, leading to a contradiction.

From the above arguments, we conclude that no virtual vertices exist in $C$. 
Since $C$ contains at least one newly added edge from $G$ to $G'$, select an arbitrary such edge $e$. 
Let the vertices connected by $e$ be $i^*$ and $i_1$, where $i^*, i_1 \in N$.
Observe that $e$ is a normal edge and that in $G$, both $i^*$ and $i_1$ have self-looping normal edges.
This implies that $\mu(i^*) \neq i^*$ and $\mu(i_1) \neq i_1$.
Define $i_2 = \mu(i_1)$. 
Following $C$ in the order $i_1, i_2$, and continuing until returning to $i_1$, we define the sequence of vertices encountered as $i_3, i_4, \dots, i_k$, where $i_k = i_1$. 
Again, let $j^*$ be the smallest index $j \geq 2$ such that the edge connecting $i_j$ and $i_{j-1}$ does not exist in $G$.
Since the edge $e$ connecting $i_k(=i_1)$ and $i_{k-1}(=i^*)$ is absent in $G$, $j^*$ must exist. 
Moreover, since the edge connecting $i_2$ and $i_1$ exists in $G$, we have $j^* \geq 3$.\par

Now, since the edge connecting $i_{j^*}$ and $i_{j^*-1}$ does not exist in $G$, it follows that in $G'$, these two vertices are connected by a normal edge and that in $G$, both $i_{j^*}$ and $i_{j^*-1}$ have self-looping normal edges.
This implies that in \( G \), the path that passes through \( i_1, i_2, \dots, i_{j^*-1} \) in the same way as \( C \) forms an alternating path, leading to a contradiction.\par

Therefore, we conclude that the original graph $G$ must contain either an alternating path or an alternating cycle.\qed\\[1ex]\par

\begin{figure}[ht]
    \centering
    \begin{tikzpicture}
        \node[circle, draw, thick] (1) at (2,2) {1};
        \node[circle, draw, thick] (3) at (0,2) {3};
        \node[circle, draw, thick] (8) at (4,2) {8};
        \node[circle, draw, thick] (4) at (2,0) {4};
        \node[circle, draw, thick] (7) at (0,0) {7};
        \node[circle, draw, thick] (2) at (4,0) {2};
        \draw[double distance=2pt] (1) -- (8);
        \draw[double distance=2pt] (1) -- (4);
        \draw[double distance=2pt] (4) -- (2);
        \draw[double=black, double distance=2pt] (3) -- (1);
        \draw[double=black, double distance=2pt] (7) -- (4);
        \draw[double=black, double distance=2pt] (8) -- (2);
    
        \node[draw, rounded corners=15pt, fit={(3) (1)}] {};
        \node[draw, rounded corners=15pt, fit={(7) (4)}] {};
        \node[draw, rounded corners=15pt, inner sep=8pt, fit={(4) (1) (2) (8)}] {};
    
    \end{tikzpicture}

    \caption{An example of biconnected component decomposition. It consists of three blocks.}
    \label{fig:biconnected component decomposition}
\end{figure}
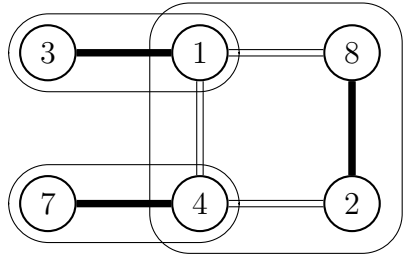

A pair of vertices is \emph{connected} if there exists a path between them. A graph \( G \) is a \emph{connected graph} if all pairs of vertices are connected. A vertex \( i \) of a graph \( G \) is a \emph{cut-vertex} if \( G \) is connected, but removing \( i \) from $G$ results in a disconnected graph. A \emph{block} is defined as a maximal subgraph that does not contain a cut-vertex. \emph{Biconnected component decomposition} involves partitioning a graph into blocks.
Figure \ref{fig:biconnected component decomposition} provides an example of biconnected component decomposition.\par

A subgraph consisting of two vertices and the single edge connecting them does not contain a cut-vertex. Thus, every edge either forms a block itself or is a subset of some block. Note that an edge cannot belong to multiple blocks. We refer to a block containing only two vertices as a \emph{trivial block}.
\par

We propose the iterative biconnected component decomposition algorithm (Algorithm \ref{alg:IBCDA}), which determines whether a matching is efficient.
\begin{algorithm}[H]
\caption{Iterative Biconnected Component Decomposition Algorithm}
\label{alg:IBCDA}
\begin{algorithmic}[1]
    \Require A matching $\mu \in \mathcal{M}$ and a preference profile $\succ\in \mathcal{P}^n$
    \Ensure Returns \texttt{true} if $\mu$ is efficient, \texttt{false} otherwise
    \If{$\mu$ contains an irrational pair}
        \State \Return \texttt{false}
    \EndIf
    \State Construct the modified graph $G'$
    \Repeat
        \State Perform a biconnected component decomposition of $G'$
        \If{All vertices (including virtual vertices) belong to a single block}
            \If{There exists a non-trivial block}
                \State \Return \texttt{false}
            \Else
                \State \Return \texttt{true}
            \EndIf
        \Else
            \For{each block $B$ and each vertex $i$ in $B$}
                \If{The special edge of $i$ is not in $B$}
                    \State Remove $i$ from $B$ (see Figure \ref{fig:remove vertices})
                \EndIf
            \EndFor
        \EndIf
    \Until{Algorithm terminates}
\end{algorithmic}
\end{algorithm}

\begin{figure}[ht]
    \centering
    \begin{tikzpicture}
        \node[circle, draw, thick] (1) at (2,2) {1};
        \node[circle, draw, thick] (3) at (0,2) {3};
        \node[circle, draw, thick] (8) at (4,2) {8};
        \node[circle, draw, thick] (4) at (2,0) {4};
        \node[circle, draw, thick] (7) at (0,0) {7};
        \node[circle, draw, thick] (2) at (4,0) {2};
        \draw[double distance=2pt] (1) -- (8);
        \draw[double distance=2pt] (1) -- (4);
        \draw[double distance=2pt] (4) -- (2);
        \draw[double=black, double distance=2pt] (3) -- (1);
        \draw[double=black, double distance=2pt] (7) -- (4);
        \draw[double=black, double distance=2pt] (8) -- (2);
    
        \node[draw, rounded corners=15pt, fit={(3) (1)}] {};
        \node[draw, rounded corners=15pt, fit={(7) (4)}] {};
        \node[draw, rounded corners=15pt, inner sep=8pt, fit={(4) (1) (2) (8)}] {};

        \draw[line width=3pt, ->] (5,1) -- (6,1);

        \node[circle, draw, thick] (1) at (9,2) {1};
        \node[circle, draw, thick] (3) at (7,2) {3};
        \node[circle, draw, thick] (8) at (11,2) {8};
        \node[circle, draw, thick] (4) at (9,0) {4};
        \node[circle, draw, thick] (7) at (7,0) {7};
        \node[circle, draw, thick] (2) at (11,0) {2};
        \draw[double=black, double distance=2pt] (3) -- (1);
        \draw[double=black, double distance=2pt] (7) -- (4);
        \draw[double=black, double distance=2pt] (8) -- (2);
    
        \node[draw, rounded corners=15pt, fit={(3) (1)}] {};
        \node[draw, rounded corners=15pt, fit={(7) (4)}] {};
        \node[draw, rounded corners=15pt, fit={(2) (8)}] {};
    
    \end{tikzpicture}

    \caption{This represents the case where vertices 1 and 4 are removed from the block containing vertices 1, 4, 2, and 8.}
    \label{fig:remove vertices}
\end{figure}
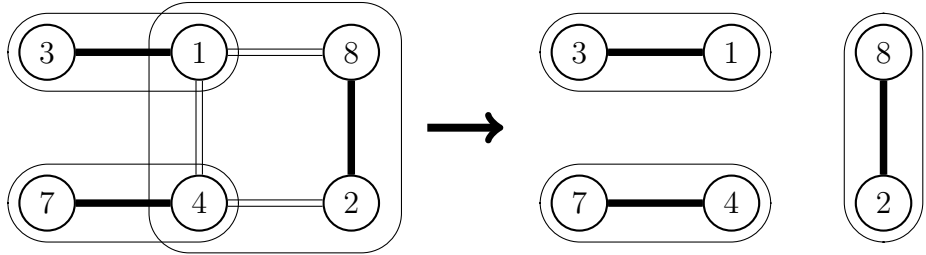

Note that Algorithm \ref{alg:IBCDA} always terminates in a finite number of steps.
If $G'$ contains an alternating cycle, then at the termination of Algorithm \ref{alg:IBCDA}, there exists a block containing that alternating cycle, which is a non-trivial block.
On the other hand, the following proposition also holds.

\prop\label{prop:morrill}
At the termination of Algorithm \ref{alg:IBCDA}, if a non-trivial block exists, then the graph $G'$ contains an alternating cycle.\par
\proof
The proof follows the same reasoning as Proposition 1 in \cite{morrill2010roommates}.
\qed\\[1ex]\par

From Lemma \ref{lem:graph and efficiency}, Lemma \ref{lem:graph and modified graph}, and Proposition \ref{prop:morrill}, we obtain the following theorem.

\thm
The iterative biconnected component decomposition algorithm determines whether a matching satisfies efficiency.
\\[1ex]\par

Moreover, it runs in $O(n^2)$ time.\footnote{See Appendix A.1 of \cite{morrill2010roommates}.}


\printbibliography

\end{document}